%% file: conference_041818.tex
\documentclass[conference]{IEEEtran}
\IEEEoverridecommandlockouts
\usepackage{cite}
\usepackage{amsmath,amssymb,amsfonts}
\usepackage{algorithmic}
\usepackage{graphicx}
\usepackage{textcomp}
\usepackage{xcolor}
\usepackage{url}
\def\BibTeX{{\rm B\kern-.05em{\sc i\kern-.025em b}\kern-.08em
    T\kern-.1667em\lower.7ex\hbox{E}\kern-.125emX}}

\DeclareMathOperator*{\argmin}{arg\,min}

\begin{document}

\title{Smart Contract-based Computing Resources Trading in Edge Computing
}

\author{
    \IEEEauthorblockN{
        Jinyue Song\IEEEauthorrefmark{1},
        Tianbo Gu\IEEEauthorrefmark{1},
        Yunjie Ge \IEEEauthorrefmark{2},
        Prasant Mohapatra\IEEEauthorrefmark{1}
    }
    \IEEEauthorblockA{
        \IEEEauthorrefmark{1}
        Department of Computer Science, University of California, Davis, CA, USA\\
        \IEEEauthorrefmark{2}
        University of San Francisco, USA\\
        Email: \{jysong,tbgu,pmohapatra\}@ucdavis.edu, yge7@dons.usfca.edu	
    }
}

\maketitle

\input{abstract.tex}
\input{intro.tex}

\input{problem_formulation}
\input{model.tex}
\input{eval.tex}
\input{related.tex}
\input{conclusion.tex}

\bibliographystyle{ieeetr}
\bibliography{references}

\end{document}

%% file: abstract.tex
\begin{abstract}


In recent years, there is an emerging trend that some computing services are moving from cloud to the edge of the networks. Compared to cloud computing, edge computing can provide services with faster response, lower expense, and more security.  The massive idle computing resources closing to the edge also enhance the deployment of edge services. Instead of using cloud services from some primary providers, edge computing provides people with a great chance to actively join the market of computing resources. However, edge computing also has some critical impediments that we have to overcome. 



In this paper, we design an edge computing service platform that can receive and distribute the computing resources from the end-users in a decentralized way. Without centralized trade control, we propose a novel hierarchical smart contract-based decentralized technique to establish the trading trust among users and provide flexible smart contract interfaces to satisfy users. Our system also considers and resolves a variety of security and privacy challenges when utilizing the encryption and distributed access control mechanism. We implement our system and conduct extensive experiments to show the feasibility and effectiveness of our proposed system.

\end{abstract}


%% file: intro.tex
\section{Introduction}

Cloud computing refers to massive data computing processing and analyzing through the network "cloud" consisted of multiple servers. It has some drawbacks: high latency, high operating costs for individual users, and high risks of data security and privacy issues. If storing data in the cloud, users can only trust cloud providers, such as Amazon AWS and Microsoft Azure, and rely on the protection of the data. The user loses control of the data in the cloud and does not know how data is secured.

In an edge computing network, the edge can be any functional entity from the data source to the cloud computing devices. These entities are equipped with edge computing platforms that converge network, computing, storage, and application core capabilities to provide real-time, dynamic, and intelligent computing service to end-users. Unlike processing and algorithmic decisions made in the cloud, edge computing is the action to push intelligence and computing closer, which shortens the transition distance and provides low latency computing service. Many families now have idle and powerful computing resources, such as iMac and Macbook Pro. This condition is the premise that our system can propose because users with idle computing resources can conduct computing resource trading with users who need computing services.

Even though our proposed system, based on the cloud computing and edge computing, has provided users with convenience and benefits, it still has the challenges of system operation, computing data security, and user privacy. A large number of users with idle computing resources and computing requests, make it challenging to find a suitable matching mechanism to discover each other. For example, a cloud computing network has a central controller, and all users need to be planned and arranged by it. Its performance affects the matching of all users, so the central controller is the bottleneck of system performance. In a decentralized network, such as an edge computing network, without a credit institution endorsing the user, it is difficult for the user to trust the other party and make a payment. Therefore, we have embedded blockchain\cite{blockchain} and smart contracts\cite{smartContract} to help users reach trading activities in untrusted networks. Since blockchain technology only solves the issue of user payment\cite{blockchainPayment}, it is not suitable for storing a large amount of computing data and protecting sensitive information. High latency\cite{7930225} is a disadvantage of the blockchain, so we need to design smart mechanisms to reduce the latency time and allow users to receive services faster. In addition to this, smart contracts measure operating costs by a particular gas unit, not the electricity cost from traditional programs. The execution of each code fragment consumes gas, so it requires us to design the execution code with lower complexity while balancing the requirement for reducing the latency time.

\textbf{Contribution:} In summary, our contributions break down into the following aspects:
\begin{itemize}
  \item We propose a smart contract-based edge computing system, which is controlled by smart contracts to maintain trading activities between users in the resource sharing network.
  \item We formally define this system with four layers, which can accurately reflect the data flow and interaction of each component.
  \item We propose a hierarchical smart contract group to efficiently allocate a massive influx of users into edge networks and reduce service latency time, which is explained in the Evaluation section.
  \item We provide users with flexible interfaces to choose existing smart contracts or customize a new one. It gives users the freedom to create smart contract policies and to satisfy their requirements. 
  \item We provide security service in the computation monitor layer, where service and payment are guaranteed to get delivered. We use RSA protocol to solve the data leaking problem and OneSwarm\cite{oneSwarm} to prevent unauthorized users from accessing computing data.
  \item We provide an optimal formula for system model, simulate the resource trading activities on a blockchain environment, and evaluate the system performance based on different allocation smart contracts design. 
\end{itemize}

\textbf{Roadmap:} The rest of the paper is organized in the following sections: Section 2 describes the edge computing system mechanism, including the overview architecture, participants in the network, and the formalized essential variables. Also, we explore the optimal solution in latency and operating costs in our design. Section 3 presents the system design with system components in each layer and how the layers interact with each other. Section 4 shows our simulation and evaluation. Finally, in section 5 and 6, we give the paper related work and conclusion.

%% file: problem_formulation.tex
\section{System Model}
In our proposal, the main participants of the system are consumers seeking computing resources, service providers selling computing resources, and a mechanism managing users' resource trading activities. Consumers and providers carry out resource trading with the technical support of blockchain. Under the premise of reasonable cost operation, our system should give users a platform that can match them efficiently and with low latency, and also provide a secure data transmission solution.



\subsection{Computation Consumer}
Consumers firstly register with the system and submit their conditions, including budget, CPU, bandwidth, and location. Then, based on the provided regions, these consumers enter the local edge computing network to match the service providers. Consumers pay for the computation resource provided by the service providers, and the system will monitor and verify their trading activities. 
We consider that each consumer instance is a distinct representative in the system. If there is a request from this consumer, we consider there will be a different consumer instance entering the system.

\subsection{Service Provider}
    \begin{figure}
    \centering
      \includegraphics[width=1 \columnwidth]{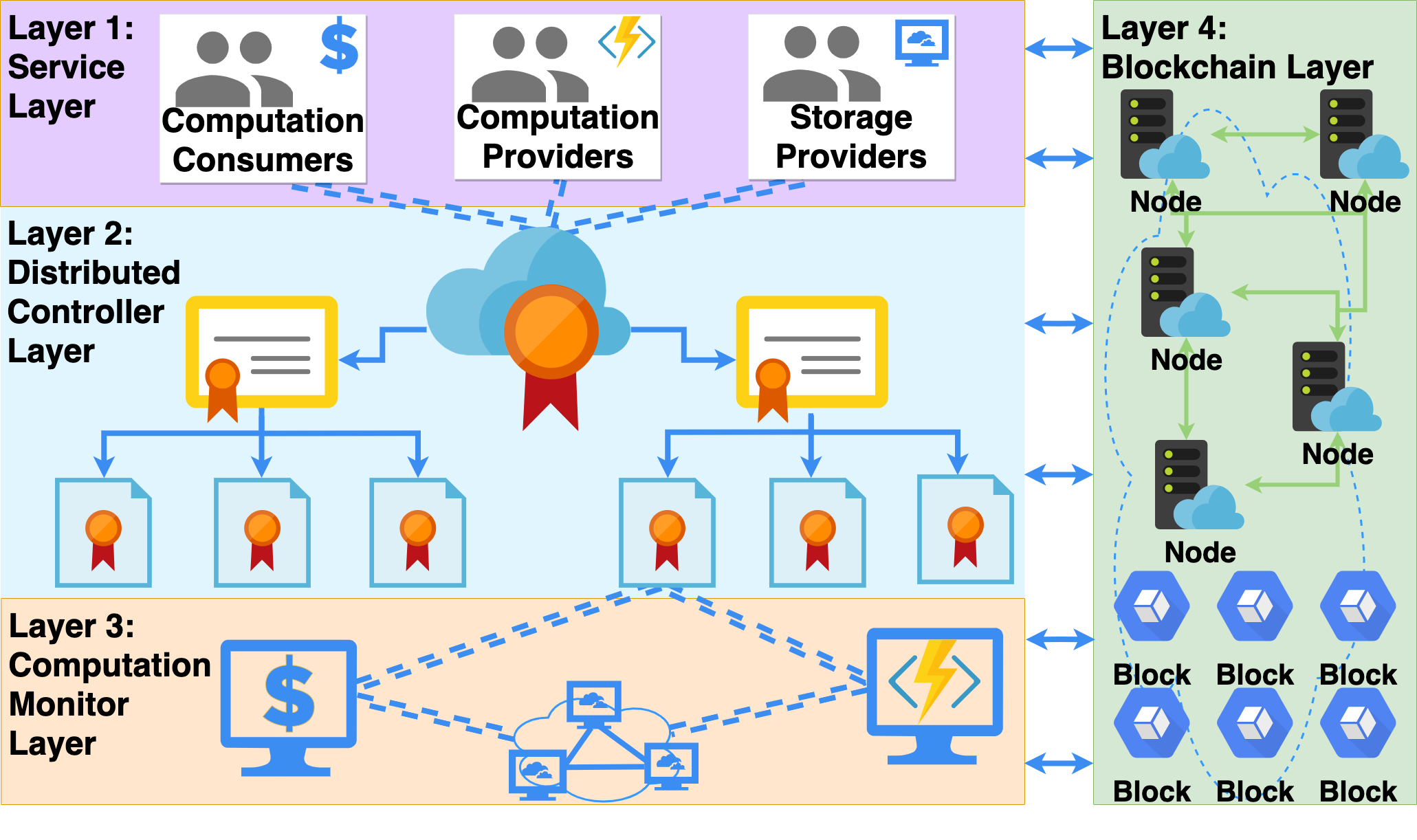}
      \caption{Smart Contract-based Computing Resources Trading Architecture.}~\label{fig:figure1}
    \end{figure}
There are two types of service providers: the computing provider and the storage provider, who provide the consumers with the functions of computing and storage in their local machines. Our proposed system will precheck providers' conditions and match qualified ones with the consumer. Then, the qualified computation provider supports the consumer for his computation tasks with his machines. This process is monitored by our system, which is responsible for managing the remuneration and working schedule. 
Similarly, the storage providers provide temporary storage services on their machines for consumers. We embed a distributed storage mechanism with encryption in our system to guarantee the data security. Storage providers can store the distributed data only without reading or writing permission. Also, the entire data distribution status is invisible to storage providers. So, it is impossible to hack the local data chunk because the storage provider does not have keys to decrypt it. Therefore, the security of all data is guaranteed.

\subsection{Matching and Resource Trading}
We propose a hierarchical structure in the system to support users' resource trading activity. This structure assumes the responsibilities of user registration in the first layer, allocation and status verification in the second layer, and trading activity supervision in the third layer. Resource trading is our primary purpose. Our design has the following three advantages: reduce the matching time of users, increase their engagement success rate, and optimize the minimum operating cost.

Users, including consumers and service providers, register to the service layer first. 
Then, in the second layer, our system will allocate users to the corresponding regional edge network according to the location provided at the time of registration, regardless of user types. After that, in the third layer, consumers distribute computing data to storage providers and do computation tasks remotely on provider's machines. The consumer and service providers trade directly under the supervision of our system at layer three. Our proposed system collects consumer's deposits and then sends a commission to the service providers according to service time. This process ensures the fairness of the trading and maximizes the revenue of the service providers. Other than that, this process provides users with multiple trading options to meet their various requirements. For example, some consumers focus on faster computing services, and some service providers want higher profit returns. They can choose a trading option that matches them with suitable partners to meets the requirements.

\subsection{Latency Minimization}
Latency is the time difference between the $t_i$ when the consumer enters the edge computing network and $t_{j}$ when he gets matched with the service provider. We consider that latency is affected by four factors: 1. the consumer's willingness to pay $bid^C$ for the computing service, 2. service providers' charging price $charge^P$, 3. the number of consumers' satisfied requirements, and 4. the waiting queue length $|Contract^m.Queue|$ in the matchmaker smart contract, which will be defined and explained detailedly in the System Design section. We extract dominating factors 1 and 2 for the formula. After registering in the distributed controller smart contract in layer one, the consumer will be allocated to the corresponding local edge computing network and then into the queue of the matchmaker smart contract, which finds service providers to engage this consumer. We thus establish the latency formula:

\begin{eqnarray}
Latency = \{bid^C, charge^P, Cond^C_P,\nonumber \\|Contract^m.Queue|, \delta\}
\end{eqnarray}
where we have a transition function $\delta$ provided by the matchmaker to determine the value of the latency for consumer $C$, $latency^C = \delta(bid^C, charge^P, Cond^C_P, |Contract^m.Queue|)$,  and the conditions between the consumer and service providers are defined as $Cond^C_P=\{Cond_1, Cond_2, ..., Cond_n | Cond_i = \{1, 0\}\}$. In our design, when the consumer's bid is greater than or equal to the service provider's charging price, the two parties will further compare the hardware conditions and reach an engagement. Otherwise, this consumer will be pushed to the $Contract^m$'s waiting queue for incoming providers. In our default engagement algorithm, n number of service providers are sorted in the binary search tree structure, and the searching time complexity is $O(lg(n))$. For m number of consumers, the total consumption time is $O(m \times lg(n))$ on average.

\subsection{Engagement Maximization}
The matchmaker smart contract manages the engagement activity between the consumer and service providers in layer two. We use the binary search tree instead of a linked list or hash table to store service providers because its time complexity is smaller than the other two. The time costs of the insertion, deletion, and lookup in the tree structure is $O(lg(n))$, but the other two structures implemented by Solidity have $O(n)$. Then, both consumers and service providers could select the smart contract, Min Latency, for example, which could maximize their benefit. In the next section, we will present that the engaged consumer and service providers collaborate with the monitor smart contract. 

The engagement rate is the number of matched consumers divided by the total number of all consumers. The role of our proposed system is to efficiently match consumers and service providers in the edge computing network. Engagement rate \textbf{EngRate} is a good measure of system performance, which is defined as follows:
\begin{eqnarray}
EngRate = \frac{\sum_1^{|C|}\gamma(\{Cond^{C_i}_{P_i})}{|C|}
\end{eqnarray}
for all $1 \leq i \leq |C|$ and $|C|$ is the total number of consumers entering this network. In the formula, we define $\gamma$ is a function provided by the matchmaker smart contract to determine if these consumer and service providers could be engaged based on their conditions. It returns binary values: 1 for successful engagement and 0 for failure. 

\subsection{Minimum Operating Cost Optimization}
At the system level, we are concerned about the operating costs of the entire system, including five factors: adding new users, deleting invalid users, retrieving and matching users, regular communication between all parties, and building smart contracts. The cost of building a smart contract is a fixed constant, but the value of the other four factors will increase with a larger number of users, no matter the implemented data structure. We want to slow down the cost growth as the number of users increases, to ensure that smart contracts can operate within a controllable cost range. At the same time, we want to find the optimal five parameters for the operating cost formula in polynomial time complexity, which turns into a P vs. NP problem. Thus, we have the cost formula:
\begin{eqnarray}
Cost = \{Op^*, Eng, Comm, Setup, \lambda\}
\end{eqnarray}
where $Op^*$ represents for all the combination of addition, deletion and lookup operations, $Eng$ is the engagement process, which can be calculated by $EngRate \times |C|$, $Setup$ is the essential cost to initialize the contract, and $\lambda$ is the transition function to calculate the cost. 

We calculate the average value and variance of the operating cost on each pair of consumer and service providers. The purpose is to get a stable overhead at the minimum operating cost for optimal parameters. First, we get the arguments that allow variance and average of the total cost to be minimum:
\begin{eqnarray}
args_{var} = \argmin var(\lambda^{C,P})\\
args_{avg} = \argmin avg(\lambda^{C,P})
\end{eqnarray}
where $\lambda$ calculates each of the costs for all pairs of consumers and service providers.

Then, we adjust $args_{var}$ and $args_{avg}$ into the cost function with penalty $\zeta$ in order to get the global minimum (optimal) arguments:
\begin{eqnarray}
args_{optimal} = minCost(args_{var}, args_{avg}, \zeta)
\end{eqnarray}

In the next section, we present the evaluation of our system by simulating consumers and service providers' activities in a private blockchain environment. 

%% file: model.tex
\section{System design}
\subsection{Overview}


Our resource sharing system contains four layers for users' engagement and computation in the edge computing network. 
As shown in Fig. 1, these four layers are Service Layer $SL$, Distributed Controller layer $DSL$, Computation Monitor layer $CML$, and Blockchain layer $BL$. 
This system supports two primary functionalities: 1. matching n number of consumers $\{C: C_1, C_2, ..., C_n\}$ and m number of service providers $\{P: P_1, P_2, ..., P_m\}$ for the win-win purpose and 2. avoiding scams during the cooperation between the two parties. The system ensures that consumers can receive qualified computing services, and service providers can receive reasonable compensation. Also, this system can guarantee the security of transmitting data and privacy of user identification. So, we introduce eight entities in this system: three types of users (consumers $C$, computation providers $P^c$, and storage providers $P^s$), three types of smart contracts (distributed controller $Contract^d$, matchmaker $Contract^m$, and intermediary $Contract^i$), nodes known as miners, and blockchain. Smart contracts are the core of this system, and they are responsible for its operation. Activities in the first three layers will be packaged into transactions and recorded on the blockchain at layer four.



    \begin{figure}
    \centering
      \includegraphics[width=1\columnwidth]{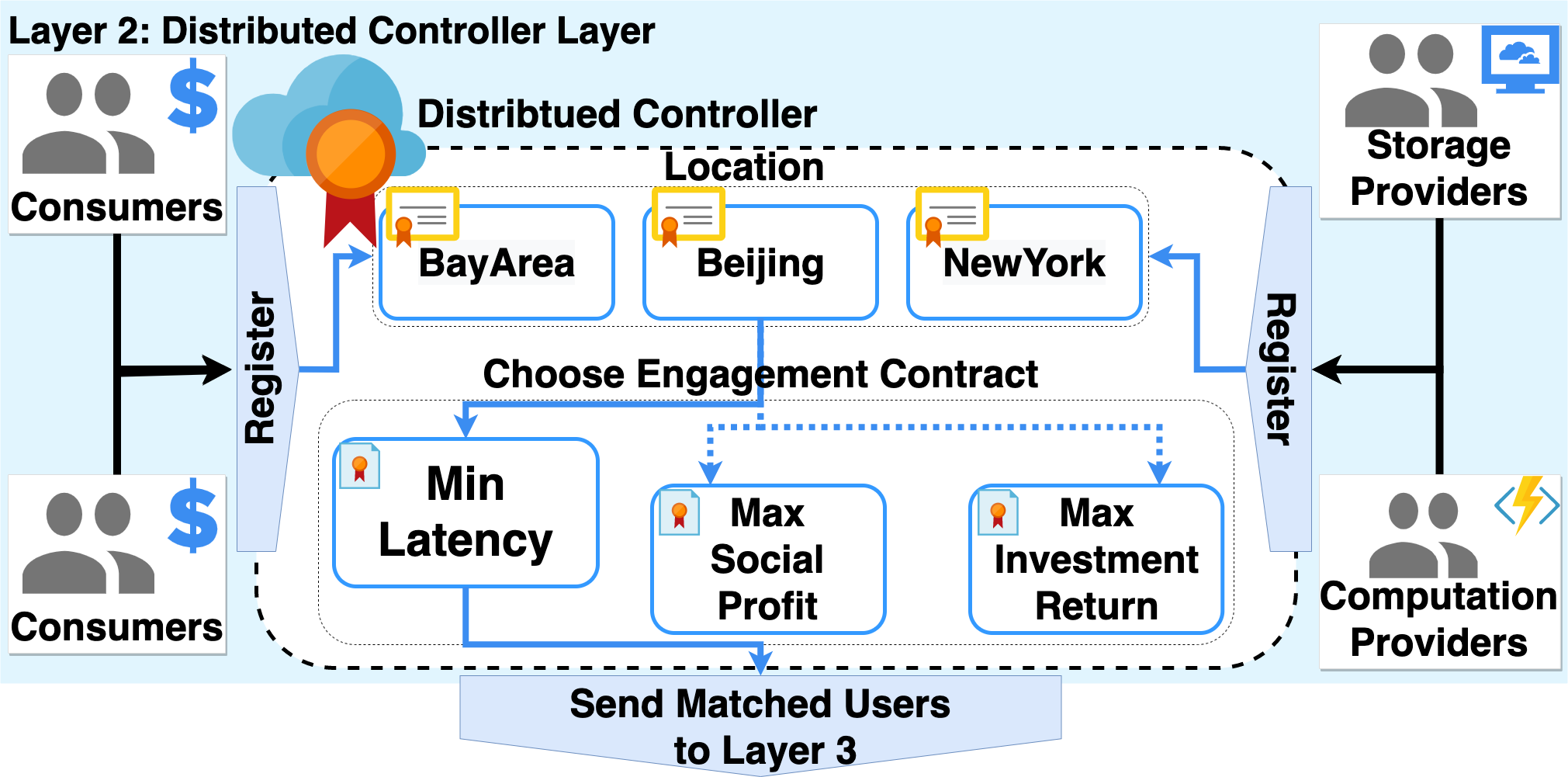}
      \caption{Layer 2 Distributed Controller Layer: illustrated by an example of consumers and providers from Bay areas, Beijing and New York.}~\label{fig:figure1}
    \end{figure}

\subsection{Service Layer}
Service layer $SL$ is the first layer in our system. It is responsible for user registration and user status collection. Users will be allocated and distributed to the local edge computing network based on their regions. Then, the second layer Distributed Controller layer will take care of users for matching mechanism. 

\textbf{Distributed Controller Contract:} In $SL$, consumers $C$ and service providers $P$ send transactions to the distributed controller $Contract^d$ first, which contains their registration information and conditions $\{Cond: Cond_1, Cond_2, ..., Cond_j\}$ like a maximum budget for the virtual machine service, and work deadline, etc. This budget is an essential factor affecting the matching result. Based on users' regional information, they are distributed to a queue of the local edge computing network by $Contract^d$ in the next layer. So, $Contract^d$ is a cross-layer contract, transferring registered users from layer one to two. Our system improves the speed of user allocation and reduces the latency of the service. Besides, users $C$ and $P$ also provide their status information, such as bandwidth, to make the system more accurately match them and improve the success rate of matching.

\subsection{Distributed Controller Layer}
Distributed Controller Layer $DCL$ is the second layer as the core joint component in this system. It accepts users from the first layer, matches consumers and service providers according to the conditions, and then passes the matched users to the third layer, allowing it to supervise the users' resource trading.

\textbf{Matchmaker Contract:} As $DCL$ shown in Fig. 2, there are three smart contracts distributed controller $Contract^d$, matchmaker $Contract^m$, and intermediary $Contract^i$, which form an inherited structure. Based on the user's regional information, $Contract^d$ assigns the users registered in layer one to their local edge computing network and passes the users to the matchmaker $Contract^m$, which is responsible for the matching of local consumers and service providers. Then, users $C$ and $P$ have the freedom to choose one of $Contract^i$s that can maximize the revenue $R$, and supervise their collaboration. For the matching mechanism, we optimize the user data storage structure of the $Contract^m$, increase its matching speed, and reduce the gas cost during the matching. Since the limitation of the gas cost measurement method in smart contracts, and more reading operations about user data than writing, we modify a tree structure to store user data. As described in section two E, the logarithmic complexity cost increment of the tree structure is much lower than the complexity of the linear complexity cost. Compared with a single central controller node, the matchmaker contract shortens the time cost of matching, which meets the user's requirements for low latency in the edge network. The evaluation section presents and quantifies our optimization of the matching mechanism. After users $C$ and $P$ match successfully, the intermediate $Contract^i$ will supervise and manage their resource trading activities in layer three.

\subsection{Computation Monitor Layer}
Computation Monitor Layer is a specific execution layer for user resource trading. It receives matched users from the second layer and supervises their resource trading activities, ensuring that the trading activities are fair for both sides.

\textbf{Intermediary Contract:} Shown in Fig. 3, the intermediary $Contract^i$ inherits the paired consumer and service providers from the matchmaker smart contract in layer two. We build a security transmission channel and payment mechanism between consumers and service providers. Our proposed transmission channel uses the RSA encryption mechanism to encrypt the data stored among the distributed storage providers, and embeds the OneSwarm protocol, allowing users to flexibly manage the other users' access permissions to his distributed data chunks. 
For example, the matched consumer $C_i$ processes data and computation on the Virtual Machine (VM) provided by the computation provider $P^C_j$ whose working process is monitored by $Contract^i$. Similarly, the storage providers $P^S$ provide temporary storage services on their VMs for consumers. After the computing data is encrypted and distributed among $P^S$s by the consumer $C_i$, the data access permission for $P^S_i$ is storage only. 
This design ensures that the data is secured and will not be leaked. At the same time, even if some nodes of the storage providers fail, the user can still download the entire data. Before the execution of this contract, the consumer must pre-process his data and distributively map them to storage providers, where the computation provider can access this data.

    \begin{figure}
    \centering
        \includegraphics[width=1\columnwidth]{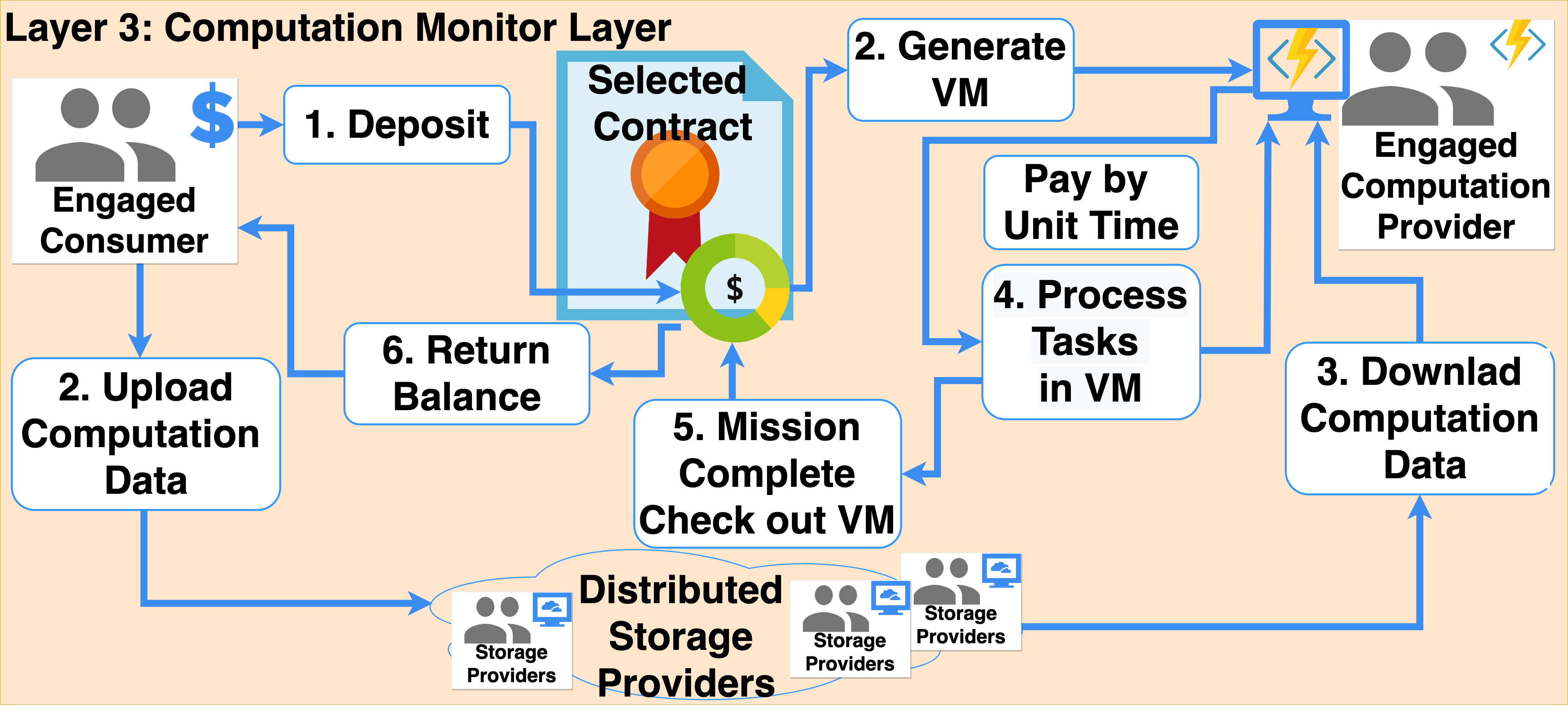}
      \caption{Layer 3 Computation Monitor Layer: selected intermediary smart contract monitors and manages matched consumer and service providers' trading activities.}~\label{fig:figure1}
    \end{figure}

There are six operations shown in Fig. 3: 
\begin{enumerate}
    \item consumer $C$ deposits his budget, and then $Contract^i$ generates the VMs in computation provider and assigns storage providers for this consumer;
    \item consumer $C$ and computation provider $P^C$ exchange their public key so that they can verify the encrypted data and signature from the other side;
    \item consumer $C$ divides the whole data into chunks and map them to tasks \emph{t}s, which have \emph{task ID}s associated with each chunk;
    \item consumer $C$ encrypts tasks by a private key and send them to storage providers $P^S$; then he updates distributed hash table (DHT), where the \emph{task ID} and \emph{storage providers IP addresses} are in pair; 
    \item consumer $C$ sends the DHT signed by his private key to computation provider $P^C$;
    \item computation provider $P^C$ verifies the DHT by the consumer $C$'s public key, and then they are ready to compute tasks for rewards in the next stage.
\end{enumerate}

OneSwarm provides data flexibility and qualified privacy. The data host can customize the access permission for other users. Assuming the consumer has two datasets, $d_0$ for provider 0 and $d_1$ for provider 1; $d_0$ is accessible for the trusted peer provider 0, not untrusted provider 1; contrary, $d_1$ is accessible for trusted provider 1, not provider 0. This design allows the consumer to manage his dataset flexibility for different computation scenarios. Since the data is available for trusted users only, unrelated users cannot read the data chunk without an assigned key. Thus, this design provides security and privacy.

$Contract^i$ plays as the intermediary between consumer $C$ and service providers $P$. It holds the consumer $C$'s deposit and guarantees service providers $P$ to be paid for every unit time $\Delta t$ when the VMs are occupied. The contract receives a notification from the consumer $C$ to check out, and then, it squares up with service providers and returns balance to the consumer. 

\subsection{Blockchain Layer}
Blockchain Layer is a data storage layer. It interacts with the other three layers and is responsible for saving all execution results, and computing data results in the system.

\textbf{Blockchain Database:} We consider the blockchain as the database for our design. As shown in Fig. 1, the data generated within layers one, two, and three, and the data transferred across layers are packaged into transactions and written to the blockchain database. This data includes user registration, matching results between consumers $C$ and service providers $P$, and progress and results of work completed under contract supervision. The computation results are stored in the blockchain because they are immutable receipts for computing consumption. Moreover, to clarify, these transactions are proofs of the computation result instead of the computation data provided by the consumers $C$. This design ensures that the data results will not be manipulated, and provides the function of verifying the computation result authenticity. The user can compare the hash values between the received computation result and the result written in the blockchain, to verify whether the current data is valid.

\subsection{Security and Privacy}
We highlight the six primary properties of this system on security and privacy aspects:
\begin{itemize}
    \item Identity isolation: participants in the network do not know each other until the smart contract matches a pair of a consumer and providers.  
    \item Data encryption: matched consumers and providers use RSA for data security.
    \item Data integrality: DHT assigns $task ID$ to service providers $IP addresses$ and records the data distribution in key-value pairs. 
    \item Access control: thanks for the protocol OneSwarm, consumer and provider could manage peers to access the data in rich options. 
    \item Trust in edge: there is no data center to filter and record all information. 
    \item Anti-fraud: due to the immutability property in the smart contract, no one can manipulate the operations and records. Participants follow the rules in code. 
\end{itemize}

%% file: eval.tex
\section{Evaluation}
We simulate the activities of users in the network of a private Ethereum blockchain and measure the performance of the edge computing system from three aspects: engagement rate, matching time, and operating cost. First, we introduce the environment configuration and program setup. Then we evaluate the system and analyze the performance data. Our purpose is to show the operability, stability, and scalability of the system.

    \begin{figure}
    \centering
      \includegraphics[width=1.0\columnwidth]{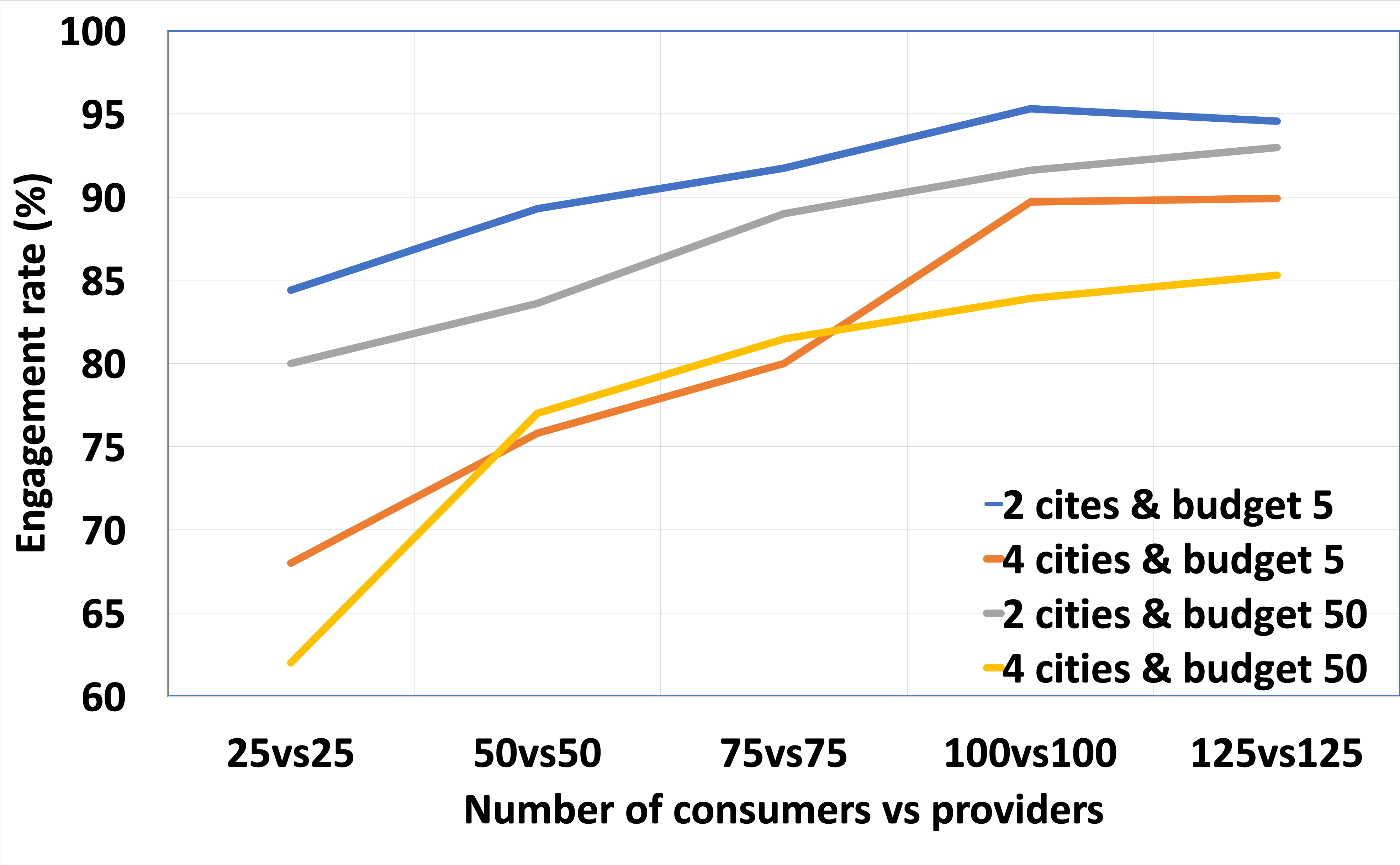}
      \caption{Average engagement rate between five different numbers of users for four condition combinations.}~\label{fig:figure1}
    \end{figure}

\subsection{System Implementation}
The system runs on the macOS Version 10.15, with 2.9 GHz 6-Core Intel Core i9, 32 GB of DDR4 memory in 2400 MHz, and 8.0 GT/s Apple SSD. We simulate consumers and service providers as requests sent from a Python script to the smart contract deployed in the Ethereum private blockchain, which is generated by the software called Ganache\cite{ganache}. Our smart contracts are written in language Solidity.

In the simulation, there are the following variables: the number of users U $\in$ \{\emph{consumer C}, \emph{service provider P}\} , the distribution of the user's area l $\in$ \{${city}_0, {city}_1, ..., {city}_M$\}, the budget $budget$ $\in \{cond^C_P\}$, and the valid sustainable time $sustainable Time$ $\in \{cond^C_P\}$. The number of consumers and service providers $|U|$ increases from 25 to 125, with an interval of 25. The contract assigns storage providers to consumers. Their regional distribution $l$ is two or four cities respectively; the budgets $cond_{budget}$ are \$5 or \$50, respectively, and randomly increase in the range of 0 to 1; the sustainable time $cond_{sustainable Time}$ of the two types of users coincides with at least three quarters, and the difference is controlled by the stochastic equation $random()$. In order to get accurate and bias-free results, each combination is looped for ten times to calculate their average $avg$ and variance $var$ values.

\subsection{Performance Evaluation}
    \begin{figure}
    \centering
      \includegraphics[width=1.0\columnwidth]{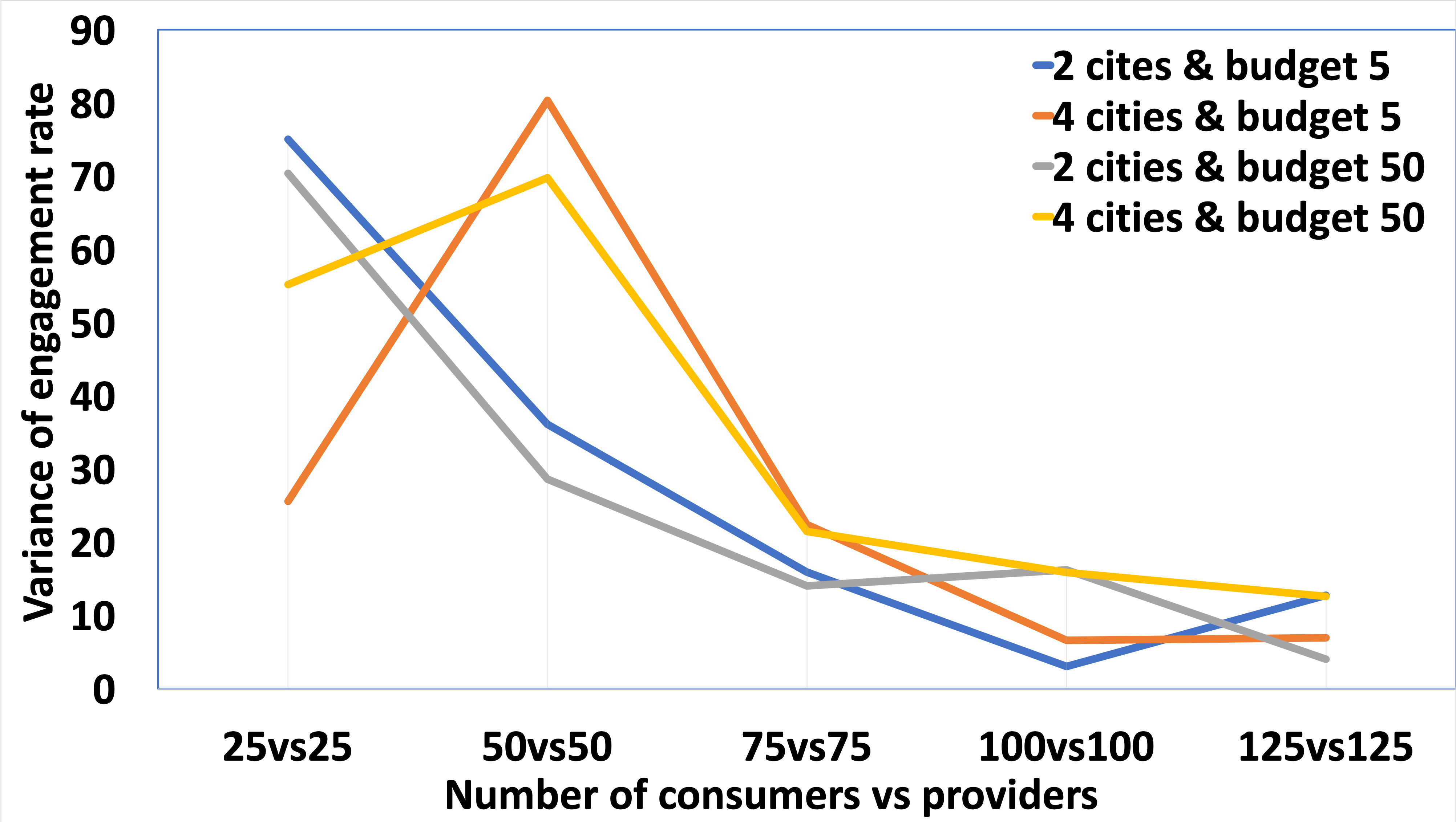}
      \caption{Variance of average engagement rate for ten-round simulations.}~\label{fig:figure1}
    \end{figure}

\subsubsection{Engagement Rate Evaluation}
We measured the engagement rate in set $ER_{set} = \{EngRate^{C_i}_{P_i}\}$ of consumers $C$ and service providers $P$. It is expected that the system can still have a high engagement rate when the number of users $|U|$ increases, and the matching conditions $cond$s become more complicated. Fig. 4 shows that the engagement ratios of the four combinations increase when the user number increases in all conditions. Fig. 5 shows that the fluctuation of the engagement rate decreases as the number of users increases. 25vs25 shows to be an outlier, because the sample size is too small to simulate the enriched service providers with different conditions, and it can not satisfy consumers. So, in Fig. 4, the engagement rate is relatively low; in Fig. 5, engagement rates are all similarly low, leading to the low variance value; in Fig. 6, the small sample size and the lacked diversity cause few users' conditions can be matched so that the system simulation can terminate earlier. 
Based on these three graphs, we can conclude that when there are more users, the engagement rate and the system performance become higher and more stable. Thus, this system is highly scalable.

\subsubsection{Matching Time Cost Evaluation}
    \begin{figure}
    \centering
      \includegraphics[width=1.0\columnwidth]{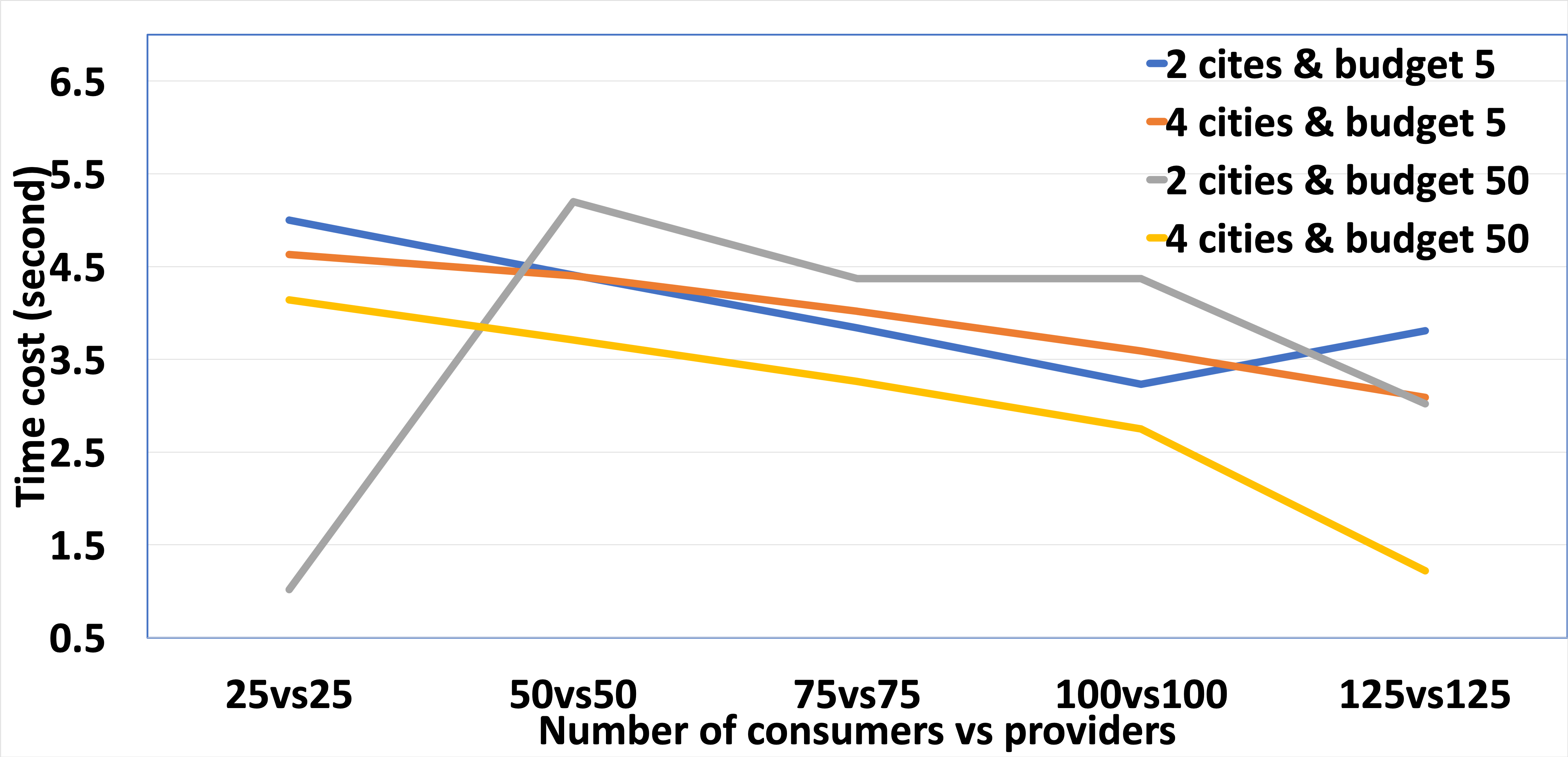}
      \caption{Average time cost for one engagement.}~\label{fig:figure1}
    \end{figure}
    
    \begin{figure}
    \centering
      \includegraphics[width=1.0\columnwidth]{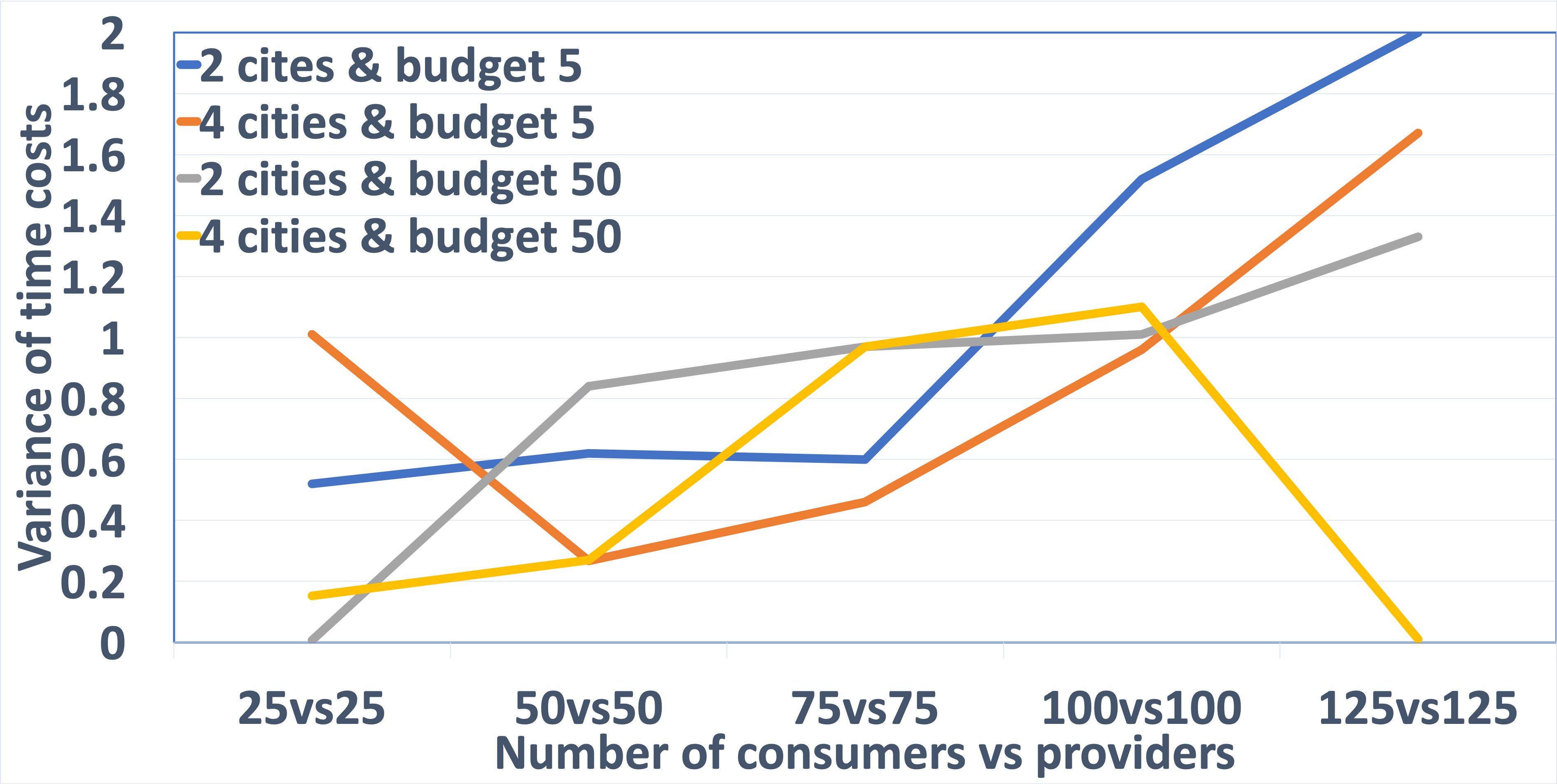}
      \caption{Variance of time cost for a engagement among ten-round simulations.}~\label{fig:figure1}
    \end{figure}
    
Regarding matching time $Match = \{match_P^C\}$ for all consumers $C$ and service providers $P$, we care about its average and variance values: $avg(match_P^C)$ and $var(match_P^C)$. Fig. 6 shows that when users have more budgets and are distributed into more areas like four cities, the average matching time will decrease. We can predict that when the network scale expands to cover many users, cities, and smart contracts, the time cost will be reduced more, and the latency time will be further reduced. However, we can see that in Fig. 7, the time cost is not perfectly stable, and its variance increases when we have more users in the system. 
    
\subsubsection{Minimum Operating Cost Evaluation}
    \begin{figure}1
    \centering
      \includegraphics[width=1.0\columnwidth]{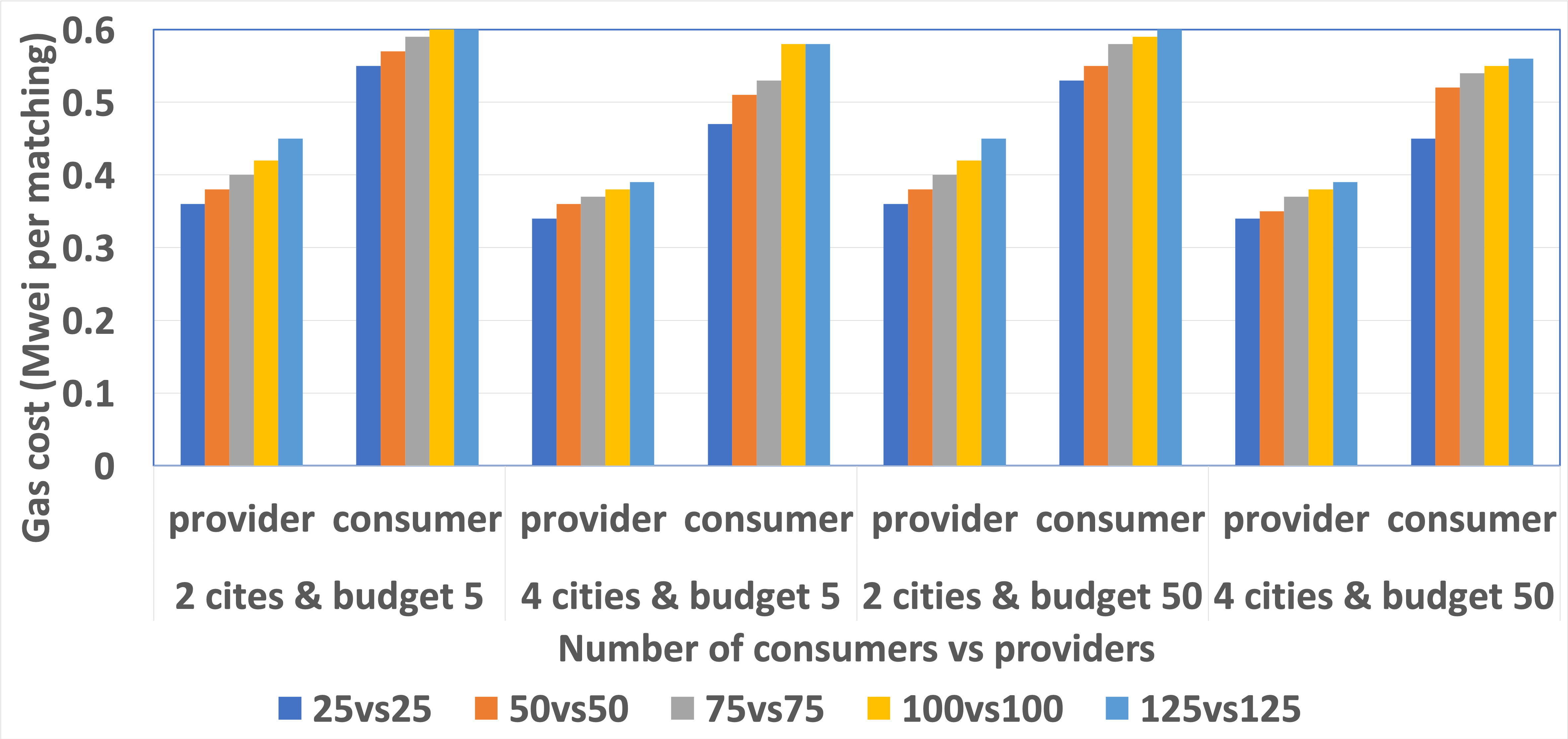}
      \caption{Average gas cost for one pair of consumer and providers.}~\label{fig:figure1}
    \end{figure}
    
    \begin{figure}
    \centering
      \includegraphics[width=1.0\columnwidth]{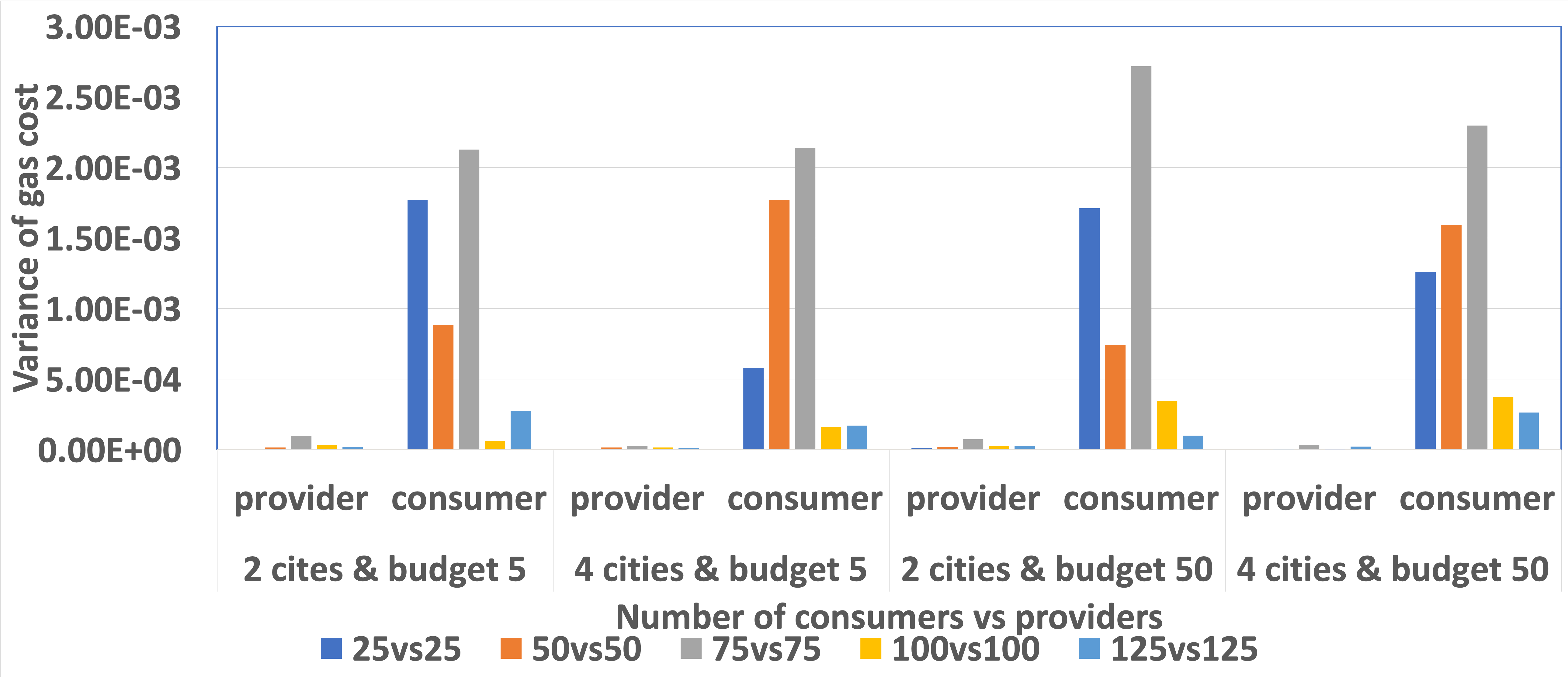}
      \caption{Variance of gas cost in matching consumer and provider.}~\label{fig:figure1}
    \end{figure}
    
Operating cost is the amount of gas used for contract executions. In Fig. 8, the average cost $avg({cost}_{C, P}^{contract})$ of providers and consumers is a stable and linear increase. This linear relationship is acceptable because as the number of users increases, the searching workload for matching will increase. This linear relationship can be represented as $avg({cost}_{C, P}^{contract}) = \alpha_{C, P} \times |C+P| + \beta_{C, P} + \epsilon_{C, P}$.

Fig. 9 shows that the variance of consumers' gas cost $var({cost}_{C, P}^{contract})$ is much greater than providers'. Because in our simulation, a consumer will traverse through all providers on the waiting list to find the matching target. Even if the consumers' cost variance is much larger than that of the providers, the absolute value of consumers' cost variance is almost negligible. Therefore, we believe that the operating cost is very stable and can be added to the system model as linear variables as $\sigma_{C, P} \times {cost}_{C, P}^{contract}$.

%% file: related.tex
\section{Related Work}
In the data security aspect, the RSA encryption mechanism allows our edge computing system to deliver data to the trusted and untrusted ends\cite{10.1007/978-3-540-24676-3_30}. However, this solution cannot fully solve our data storage and security challenges.
A distributed storage system called Bigtable\cite{Bigtable} shows that It can store the structured data among severs for low-latency and high throughput. However, this method cannot handle data in inconsistent formats.

From the edge computing view, previous work focuses more on the mobile\cite{edgeIoT} computation or the computation in the 5G network\cite{computingIn5G}, which has a faster response and high-performance. However, those mobile devices are low in sustainable energy, and computing power cannot satisfy the advanced requirements, like video processing. 
Also, the traditional cloud computing system could be inaccessible or unstable when its central server or cluster is hacked\cite{cloudDrawbacks}. 

The next aspect comes to the blockchain and smart contract. The decentralized system \cite{Hawk} could execute the same smart contracts asynchronously and have a consistency in data recording. This design inspires our work that the edge computing system could be decentralized, deployed on the blockchain, and executed by smart contracts, which solves the central controller drawbacks and potential crash issues\cite{dataIssue}.

%% file: conclusion.tex
\section{Conclusion}
In this article, we design this blockchain-enabled computing system to coordinate consumers and providers for resource trading. This system uses the blockchain as a database and smart contracts as intermediaries to automatically execute designed programs without being manipulated. This system provides data security by using RSA public and private keys to encrypt tasks and sign messages. Because of the distributed storage and OneSwarms protocols, this system allows consumers to flexibly manage providers' permission about the data access in untrusted networks. This system achieves the dual purpose of flexible data management and privacy and security protection. Except for these challenges, our system resolves the user matching challenge, operating cost challenge, and smart contract inheritance challenge when implementing flexible contract interface, customized tree structure, and four-layer structure. 